\UseRawInputEncoding
\documentclass[%
 reprint,
superscriptaddress,
 amsmath,amssymb,
 aps,
]{revtex4-2}

\usepackage{graphicx}
\usepackage{dcolumn}
\usepackage[usenames,dvipsnames,svgnames,table]{xcolor}	
\usepackage[normalem]{ulem}
\usepackage{color}
\usepackage{bm}
\RequirePackage{graphicx}
\RequirePackage{color}
\usepackage{mathtools}
\usepackage{extarrows}
\usepackage{bm}
\RequirePackage{latexsym}
\RequirePackage{xspace}

\newcommand{\meas}{(7.07~\pm~0.02~\text{(stat.)}~\pm~0.11~\text{(syst.)})~\times~10^{18}}
\newcommand{\ximeas}{0.45~\pm 0.03~\text{(stat.)} \ \pm 0.05 \ \text{(syst.)}}

\newcommand{\gAMeasSyst}{1.11~\pm 0.03~\text{(stat.)} \ \pm 0.05 \ \text{(syst.)}}

\usepackage{amsmath,amsfonts,bm}
\usepackage{wasysym}
\usepackage{color}
\usepackage{lineno}
\usepackage{appendix}

\begin{document}

\preprint{APS/123-QED}

\title{ Measurement of the $2\nu\beta\beta$ decay rate and spectral shape of $^{100}$Mo from the CUPID-Mo experiment}

\collaboration{CUPID-Mo collaboration}

\author{C.~Augier }
\affiliation{Univ Lyon, Universit\'{e} Lyon 1, CNRS/IN2P3, IP2I-Lyon, F-69622, Villeurbanne, France }

\author{A.~S.~Barabash }
\affiliation{National Research Centre ``Kurchatov Institute", Kurchatov Complex of Theoretical and Experimental Physics, 117218 Moscow, Russia }

\author{F.~Bellini }
\affiliation{Dipartimento di Fisica, Sapienza Universit\`a di Roma, P.le Aldo Moro 2, I-00185, Rome, Italy }
\affiliation{INFN, Sezione di Roma, P.le Aldo Moro 2, I-00185, Rome, Italy}

\author{G.~Benato }
\affiliation{INFN, Laboratori Nazionali del Gran Sasso, I-67100 Assergi (AQ), Italy }
\affiliation{INFN, Gran Sasso Science Institute, I-67100 L'Aquila, Italy}

\author{M.~Beretta }
\affiliation{ University of California, Berkeley, California 94720, USA }

\author{L.~Berg\'e }
\affiliation{Universit\'e Paris-Saclay, CNRS/IN2P3, IJCLab, 91405 Orsay, France }

\author{J.~Billard }
\affiliation{Univ Lyon, Universit\'{e} Lyon 1, CNRS/IN2P3, IP2I-Lyon, F-69622, Villeurbanne, France }

\author{Yu.~A.~Borovlev }
\affiliation{Nikolaev Institute of Inorganic Chemistry, 630090 Novosibirsk, Russia }

\author{L.~Cardani }
\affiliation{INFN, Sezione di Roma, P.le Aldo Moro 2, I-00185, Rome, Italy}

\author{N.~Casali }
\affiliation{INFN, Sezione di Roma, P.le Aldo Moro 2, I-00185, Rome, Italy}

\author{A.~Cazes }
\affiliation{Univ Lyon, Universit\'{e} Lyon 1, CNRS/IN2P3, IP2I-Lyon, F-69622, Villeurbanne, France }

\author{E. ~Celi}
\affiliation{INFN, Laboratori Nazionali del Gran Sasso, I-67100 Assergi (AQ), Italy }
\affiliation{INFN, Gran Sasso Science Institute, I-67100 L'Aquila, Italy}

\author{M.~Chapellier }
\affiliation{Universit\'e Paris-Saclay, CNRS/IN2P3, IJCLab, 91405 Orsay, France }

\author{D.~Chiesa}
\affiliation{Dipartimento di Fisica, Universit\`{a} di Milano-Bicocca, I-20126 Milano, Italy }
\affiliation{INFN, Sezione di Milano-Bicocca, I-20126 Milano, Italy}

\author{I.~Dafinei }
\affiliation{INFN, Sezione di Roma, P.le Aldo Moro 2, I-00185, Rome, Italy}

\author{F.~A.~Danevich }
\affiliation{Institute for Nuclear Research of NASU, 03028 Kyiv, Ukraine }
\affiliation{INFN, Sezione di Roma Tor Vergata, Via della Ricerca Scientifica 1, I-00133, Rome, Italy}

\author{M.~De~Jesus }
\affiliation{Univ Lyon, Universit\'{e} Lyon 1, CNRS/IN2P3, IP2I-Lyon, F-69622, Villeurbanne, France }

\author{T.~Dixon}\email{toby.dixon@ijclab.in2p3.fr}
\affiliation{Universit\'e Paris-Saclay, CNRS/IN2P3, IJCLab, 91405 Orsay, France }
\affiliation{IRFU, CEA, Universit\'{e} Paris-Saclay, F-91191 Gif-sur-Yvette, France }
\author{L.~Dumoulin }
\affiliation{Universit\'e Paris-Saclay, CNRS/IN2P3, IJCLab, 91405 Orsay, France }

\author{K.~Eitel }
\affiliation{Karlsruhe Institute of Technology, Institute for Astroparticle Physics, 76021 Karlsruhe, Germany }

\author{F.~Ferri }
\affiliation{IRFU, CEA, Universit\'{e} Paris-Saclay, F-91191 Gif-sur-Yvette, France }

\author{B.~K.~Fujikawa }
\affiliation{ Lawrence Berkeley National Laboratory, Berkeley, California 94720, USA }

\author{J.~Gascon }
\affiliation{Univ Lyon, Universit\'{e} Lyon 1, CNRS/IN2P3, IP2I-Lyon, F-69622, Villeurbanne, France }

\author{L.~Gironi }
\affiliation{Dipartimento di Fisica, Universit\`{a} di Milano-Bicocca, I-20126 Milano, Italy }
\affiliation{INFN, Sezione di Milano-Bicocca, I-20126 Milano, Italy}

\author{A.~Giuliani} 
\affiliation{Universit\'e Paris-Saclay, CNRS/IN2P3, IJCLab, 91405 Orsay, France }

\author{V.~D.~Grigorieva }
\affiliation{Nikolaev Institute of Inorganic Chemistry, 630090 Novosibirsk, Russia }

\author{M.~Gros }
\affiliation{IRFU, CEA, Universit\'{e} Paris-Saclay, F-91191 Gif-sur-Yvette, France }

\author{D.~L.~Helis }
\affiliation{IRFU, CEA, Universit\'{e} Paris-Saclay, F-91191 Gif-sur-Yvette, France }
\affiliation{INFN, Laboratori Nazionali del Gran Sasso, I-67100 Assergi (AQ), Italy }

\author{H.~Z.~Huang }
\affiliation{Key Laboratory of Nuclear Physics and Ion-beam Application (MOE), Fudan University, Shanghai 200433, PR China }

\author{R.~Huang }
\affiliation{ University of California, Berkeley, California 94720, USA }

\author{L.~Imbert}
\affiliation{Universit\'e Paris-Saclay, CNRS/IN2P3, IJCLab, 91405 Orsay, France }

\author{J.~Johnston }
\affiliation{Massachusetts Institute of Technology, Cambridge, MA 02139, USA }

\author{A.~Juillard }
\affiliation{Univ Lyon, Universit\'{e} Lyon 1, CNRS/IN2P3, IP2I-Lyon, F-69622, Villeurbanne, France }

\author{H.~Khalife }
\affiliation{Universit\'e Paris-Saclay, CNRS/IN2P3, IJCLab, 91405 Orsay, France }

\author{M.~Kleifges }
\affiliation{Karlsruhe Institute of Technology, Institute for Data Processing and Electronics, 76021 Karlsruhe, Germany }

\author{V.~V.~Kobychev }
\affiliation{Institute for Nuclear Research of NASU, 03028 Kyiv, Ukraine }

\author{Yu.~G.~Kolomensky }
\affiliation{ University of California, Berkeley, California 94720, USA }
\affiliation{ Lawrence Berkeley National Laboratory, Berkeley, California 94720, USA }%

\author{S.I.~Konovalov }
\affiliation{National Research Centre Kurchatov Institute, Institute of Theoretical and Experimental Physics, 117218 Moscow, Russia }

\author{J.~Kotila}
\affiliation{Department of Physics, University of Jyv\"{a}skyl\"{a}, PO Box 35, FI-40014, Jyv\"{a}skyl\"{a}, Finland }
\affiliation{Finnish Institute for Educational Research, University of Jyv\"{a}skyl\"{a}, P.O. Box 35, FI-40014 Jyva\"{a}skyl\"{a}, Finland }
\affiliation{Center for Theoretical Physics, Sloane Physics Laboratory, Yale University, New Haven, Connecticut 06520-8120, USA}
\author{P.~Loaiza }
\affiliation{Universit\'e Paris-Saclay, CNRS/IN2P3, IJCLab, 91405 Orsay, France }

\author{L.~Ma }
\affiliation{Key Laboratory of Nuclear Physics and Ion-beam Application (MOE), Fudan University, Shanghai 200433, PR China }

\author{E.~P.~Makarov }
\affiliation{Nikolaev Institute of Inorganic Chemistry, 630090 Novosibirsk, Russia }

\author{P.~de~Marcillac }
\affiliation{Universit\'e Paris-Saclay, CNRS/IN2P3, IJCLab, 91405 Orsay, France }

\author{R.~Mariam}
\affiliation{Universit\'e Paris-Saclay, CNRS/IN2P3, IJCLab, 91405 Orsay, France }

\author{L.~Marini }
\affiliation{ University of California, Berkeley, California 94720, USA }
\affiliation{ Lawrence Berkeley National Laboratory, Berkeley, California 94720, USA }
\affiliation{INFN, Laboratori Nazionali del Gran Sasso, I-67100 Assergi (AQ), Italy }

\author{S.~Marnieros }
\affiliation{Universit\'e Paris-Saclay, CNRS/IN2P3, IJCLab, 91405 Orsay, France }

\author{X.-F.~Navick }
\affiliation{IRFU, CEA, Universit\'{e} Paris-Saclay, F-91191 Gif-sur-Yvette, France }

\author{C.~Nones }
\affiliation{IRFU, CEA, Universit\'{e} Paris-Saclay, F-91191 Gif-sur-Yvette, France }

\author{E.~B.~Norman}
\affiliation{ University of California, Berkeley, California 94720, USA }

\author{E.~Olivieri }
\affiliation{Universit\'e Paris-Saclay, CNRS/IN2P3, IJCLab, 91405 Orsay, France }

\author{J.~L.~Ouellet }
\affiliation{Massachusetts Institute of Technology, Cambridge, MA 02139, USA }

\author{L.~Pagnanini }
\affiliation{INFN, Gran Sasso Science Institute, I-67100 L'Aquila, Italy}
\affiliation{INFN, Laboratori Nazionali del Gran Sasso, I-67100 Assergi (AQ), Italy }

\author{L.~Pattavina }
\affiliation{INFN, Laboratori Nazionali del Gran Sasso, I-67100 Assergi (AQ), Italy }
\affiliation{Physik Department, Technische Universit\"at M\"unchen, Garching D-85748, Germany }

\author{B.~Paul }
\affiliation{IRFU, CEA, Universit\'{e} Paris-Saclay, F-91191 Gif-sur-Yvette, France }

\author{M.~Pavan }
\affiliation{Dipartimento di Fisica, Universit\`{a} di Milano-Bicocca, I-20126 Milano, Italy }
\affiliation{INFN, Sezione di Milano-Bicocca, I-20126 Milano, Italy}

\author{H.~Peng }
\affiliation{Department of Modern Physics, University of Science and Technology of China, Hefei 230027, PR China }

\author{G.~Pessina }
\affiliation{INFN, Sezione di Milano-Bicocca, I-20126 Milano, Italy}

\author{S.~Pirro }
\affiliation{INFN, Laboratori Nazionali del Gran Sasso, I-67100 Assergi (AQ), Italy }

\author{D.~V.~Poda }
\affiliation{Universit\'e Paris-Saclay, CNRS/IN2P3, IJCLab, 91405 Orsay, France }

\author{O.~G.~Polischuk }
\affiliation{Institute for Nuclear Research of NASU, 03028 Kyiv, Ukraine }
\affiliation{INFN, Sezione di Roma, P.le Aldo Moro 2, I-00185, Rome, Italy}

\author{S.~Pozzi }
\affiliation{INFN, Sezione di Milano-Bicocca, I-20126 Milano, Italy}

\author{E.~Previtali }
\affiliation{Dipartimento di Fisica, Universit\`{a} di Milano-Bicocca, I-20126 Milano, Italy }
\affiliation{INFN, Sezione di Milano-Bicocca, I-20126 Milano, Italy}

\author{Th.~Redon }
\affiliation{Universit\'e Paris-Saclay, CNRS/IN2P3, IJCLab, 91405 Orsay, France }

\author{A.~Rojas }
\affiliation{LSM, Laboratoire Souterrain de Modane, 73500 Modane, France }

\author{S.~Rozov }
\affiliation{Laboratory of Nuclear Problems, JINR, 141980 Dubna, Moscow region, Russia }

\author{V.~Sanglard }
\affiliation{Univ Lyon, Universit\'{e} Lyon 1, CNRS/IN2P3, IP2I-Lyon, F-69622, Villeurbanne, France }

\author{J.~A.~Scarpaci}
\affiliation{Universit\'e Paris-Saclay, CNRS/IN2P3, IJCLab, 91405 Orsay, France }

\author{B.~Schmidt }
\affiliation{IRFU, CEA, Universit\'{e} Paris-Saclay, F-91191 Gif-sur-Yvette, France }

\author{Y.~Shen }
\affiliation{Key Laboratory of Nuclear Physics and Ion-beam Application (MOE), Fudan University, Shanghai 200433, PR China }

\author{V.~N.~Shlegel }
\affiliation{Nikolaev Institute of Inorganic Chemistry, 630090 Novosibirsk, Russia }

\author{F. ~\v{S}imkovic}
\affiliation{Faculty of Mathematics, Physics and Informatics, Comenius University in Bratislava, 842~48~Bratislava, Slovakia}
\affiliation{Institute of Experimental and Applied Physics, Czech Technical University in Prague, 128~00~Prague, Czech Republic
}
\author{V.~Singh }
\affiliation{ University of California, Berkeley, California 94720, USA }

\author{C.~Tomei }
\affiliation{INFN, Sezione di Roma, P.le Aldo Moro 2, I-00185, Rome, Italy}

\author{V.~I.~Tretyak }
\affiliation{Institute for Nuclear Research of NASU, 03028 Kyiv, Ukraine }
\affiliation{INFN, Laboratori Nazionali del Gran Sasso, I-67100 Assergi (AQ), Italy }

\author{V.~I.~Umatov }
\affiliation{National Research Centre Kurchatov Institute, Institute of Theoretical and Experimental Physics, 117218 Moscow, Russia }

\author{L.~Vagneron }
\affiliation{Univ Lyon, Universit\'{e} Lyon 1, CNRS/IN2P3, IP2I-Lyon, F-69622, Villeurbanne, France }

\author{M.~Vel\'azquez }
\affiliation{Universit\'e Grenoble Alpes, CNRS, Grenoble INP, SIMAP, 38420 Saint Martin d'H\`eres, France }

\author{B. ~Ware }
\affiliation{John de Laeter Centre for Isotope Research, GPO Box U 1987, Curtin University, Bentley, WA, Australia}

\author{B.~Welliver }
\affiliation{ University of California, Berkeley, California 94720, USA }

\author{L.~Winslow }
\affiliation{Massachusetts Institute of Technology, Cambridge, MA 02139, USA }

\author{M.~Xue }
\affiliation{Department of Modern Physics, University of Science and Technology of China, Hefei 230027, PR China }

\author{E.~Yakushev }
\affiliation{Laboratory of Nuclear Problems, JINR, 141980 Dubna, Moscow region, Russia }

\author{M.~Zarytskyy}
\affiliation{Institute for Nuclear Research of NASU, 03028 Kyiv, Ukraine }

\author{A.~S.~Zolotarova }
\affiliation{Universit\'e Paris-Saclay, CNRS/IN2P3, IJCLab, 91405 Orsay, France }

\date{\today}

\begin{abstract}
Neutrinoless double beta decay ($0\nu\beta\beta$) is a yet unobserved nuclear process which would demonstrate Lepton Number violation, a clear evidence of beyond Standard Model physics. The process two neutrino double beta decay ($2\nu\beta\beta)$ is allowed by the Standard Model and has been measured in numerous experiments.
In this letter, we report a measurement of $2\nu\beta\beta$ decay half-life of $^{100}$Mo to the ground state of $^{100}$Ru of $\meas$~yr by the CUPID-Mo experiment. With a relative precision of $\pm~1.6$ \% this is the most precise measurement to date of a $2\nu\beta\beta$ decay rate in $^{100}$Mo. In addition, we constrain higher-order corrections to the spectral shape which provides complementary nuclear structure information. We report a novel measurement of the shape factor $\xi_{3,1}=\ximeas$, which is compared to theoretical predictions for different nuclear models.
We also extract the first value for the effective axial vector coupling constant obtained from a spectral shape study of $2\nu\beta\beta$ decay.

\end{abstract}

\maketitle
  
\begin{figure}[tb]
     \centering
     \includegraphics[width=0.49\textwidth]{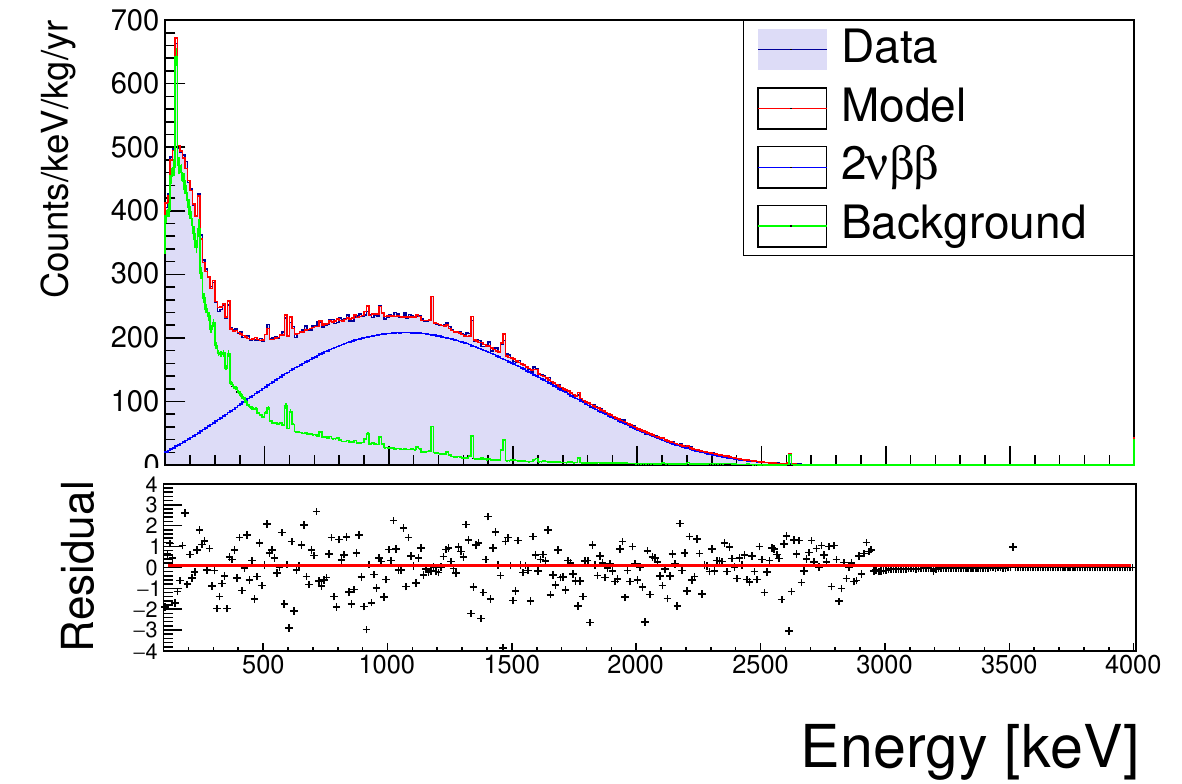}
     \caption{Fit of the $\mathcal{M}_{1,\gamma/\beta}$ spectrum showing the main contributions to the model and the residuals. The model describes well the experimental data and the spectrum above $\sim$ 500 keV is dominated by $2\nu\beta\beta$ events.}
     \label{fig:fit}
 \end{figure}

For more than 20 years it has been known that neutrinos have mass via measurements of neutrino oscillations\cite{SuperK_oscc,SNO_oscc}. This raises the question of the nature of this mass. If the neutrino is its own antiparticle, a {\it Majorana} particle, then a decay mode of some nuclei would become possible, neutrinoless double beta decay ($0\nu\beta\beta$) (see reviews \cite{0v_rev1,0v_rev2,0v_rev3}).
This decay could be observed in nuclei for which single beta decay is energetically disallowed (or disfavored by angular momentum). Two neutrons would be transformed into two protons, with the emission of only two electrons. The observation of this decay would have profound consequences for particle physics by showing that the Lepton Number is not a fundamental symmetry of nature and providing clear evidence of beyond Standard Model physics. 
\\ \indent The measurement of the decay rate could also provide a method to measure the effective neutrino mass \cite{nu_mass}. Under the light Majorana neutrino exchange mechanism the decay rate would be related to the effective Majorana mass $\langle m_{\beta\beta} \rangle$ by:
\begin{equation}
    1/T^{0\nu}_{1/2}=G_{0\nu}\cdot g_{A}^4\cdot |M_{0\nu}|^2\cdot \langle m_{\beta\beta}\rangle^2/m_e^2,
\end{equation}
where $G_{0\nu}$ is the phase space factor, $M_{0\nu}$ the nuclear matrix element (NME), $g_{A}$ the effective axial-vector coupling constant and $m_e$ the electron mass. While $G_{0\nu}$ can be calculated almost exactly \cite{kot_psf}, the NME is the result of complex many-body nuclear physics calculations (see the review \cite{NME}) and is only known to a factor of a few. To interpret the results of next-generation experiments these calculations must be improved. In addition, it has been observed that nuclear models often over-predict the decay rate of $\beta^{-}$ and $2\nu\beta\beta$. To account for this $g_A$ can be replaced with an effective value $g_{A,\text{eff}}$ \cite{Suhonen_quenching,quenching_ejiri,quenching}. Therefore there is still a possibility the $0\nu\beta\beta$ decay rate could be much lower than expected for an unrenormalized value of $g_A$ (1.27). This would have significant impact on the discovery probability of next generation experiments \cite{disc,Ettengruber:2022mtm}. To constrain this possibility new measurements are needed. 
\\ \indent Two neutrino double beta decay ($2\nu\beta\beta$) conserves Lepton Number and is allowed within the Standard Model. It has been observed in a number of nuclei \cite{Barabash_rev}. The decay rate of $2\nu\beta\beta$ decay can be described to a good approximation as:
\begin{equation}
\label{eff}
   1/T_{1/2}^{2\nu} = G_{2\nu}\cdot  g_{A}^4\cdot|M_{2\nu}|^2,
\end{equation}
where $G_{2\nu}$ is the phase space factor, $M_{2\nu}$ is the NME.
Since $0\nu\beta\beta$ and $2\nu\beta\beta$ share the same initial and final nuclear states, an accurate prediction of $T_{1/2}^{2\nu}$ and therefore an accurate description of the nuclear structure, is a necessary condition to obtain reliable estimates of $M_{0\nu}$. These measurements are often used to tune the parameters of the nuclear models. However, they cannot alone answer questions about the value of $g_{A,\text{eff}}$ since only the product $M_{2\nu}\cdot g_{A,\text{eff}}^2$ is measured. 
\\ \indent
The $2\nu\beta\beta$ decay spectrum is typically described using two approximations, the single and higher state dominance hypotheses (SSD/HSD) \cite{Simkovic:2000jm}. In these approximations, the decay is supposed to proceed via a single intermediate $1^+$ state. For the HSD model this state is an average higher energy state from the region of the Gamow-Teller resonance, while for SSD it is the lowest energy $1^+$ state.
\\ \indent The description of the $2\nu\beta\beta$ decay spectrum was improved in \cite{imp,angle}. 
In this approach, a Taylor expansion is performed in terms of the Lepton energies. The differential decay rate relates to phase space factors and NMEs as:
\begin{align}
\label{imp}
      \frac{d\Gamma}{dE}=g_{A,\text{eff}}^4|M_{GT-1}|^2\Big(&\frac{dG_0}{dE} +\xi_{3,1}\frac{dG_2}{dE}\\ \nonumber +\frac{1}{3}\xi_{3,1}^2&\frac{dG_{22}}{dE}+\big(\frac{1}{3}\xi_{3,1}^2+\xi_{5,1}\big)\frac{dG_4}{dE}\Big).
\end{align}
Here $G_0,G_2,G_{22},G_4$ are the phase space factors for different terms in the Taylor expansion. $\xi_{3,1}=M_{GT-3}/M_{GT-1}$ and $\xi_{5,1}=M_{GT-5}/M_{GT-1}$ are ratios of NMEs. By fitting the energy distribution of electrons to this model, constraints on $\xi_{3,1},\xi_{5,1}$ can be obtained which can be compared to theoretical predictions. $M_{GT-3}$ and $M_{GT-5}$ are expected to be dominated by contributions from lower-lying states due to the higher power of the energy denominators so measurement of $\xi_{3,1}$,$\xi_{5,1}$ provide complementary nuclear structure information to the half-life.
Within this model the HSD spectrum can be recovered by fixing $\xi_{3,1},\xi_{5,1}$ to zero, and the SSD approximation can be used to predict non-zero $\xi_{3,1},\xi_{5,1}$ as in \cite{imp}. $\xi$ values larger than the SSD values would indicate mutual cancellation between lower and higher lying states.
\\ \indent As described in \cite{imp}, a measurement of $\xi_{3,1}$ and the half-life can be used to extract a value for $g_{A,\text{eff}}$ of:
\begin{equation}
\label{gA_form}
    g_{A,\text{eff}}^4=\frac{T_{1/2}^{-1}\times \xi_{3,1}^2}{M_{GT-3}^2G},
\end{equation}
where $G=G_0+\xi_{3,1}G_2+\xi_{3,1}^2G_{22}/3+(\xi_{3,1}^2/3+\xi_{5,1})G_4$. $M_{GT-3}$ can be computed reliably within the *interacting shell model (ISM) which describes accurately low lying states of nuclei.
\\ \indent
So far the analysis to extract the $\xi$ factors has only been performed by the KamLAND-Zen experiment \cite{KamLAND-Zen:2019imh} which established an upper bound on $\xi_{3,1}$ in $^{136}$Xe decays, which is still compatible with both the ISM and pn-QRPA calculations. 
\\ \indent
We compute $2\nu\beta\beta$ NMEs $M_{GT-1}$, $M_{GT-3}$, and $M_{GT-5}$ within the proton-neutron quasi-particle random-phase approximation (pn-QRPA) \cite{Simkovic:2013qiy} and described in more detail in Appendix \ref{theory}. These calculations are performed for a range of $g_{A,\text{eff}}$ values. We compute the phase space factors $G_0,G_2,G_4,G_{22}$ considering Dirac wave functions with finite nuclear size and electron screening as in \cite{kot_psf}.
\\ \indent The experimental signature of $2\nu\beta\beta$ decay is a continuous spectrum in the summed energies of the electrons. Differentiation of the signal from background is more challenging than for $0\nu\beta\beta$ decay: the decay rate must be extracted from a fit to the full spectrum using detailed simulations of the various contributions to the experimental background (see for example \cite{CUPID_0_bkg,CUORE_2vbb,lumineu_2vbb,GERDA_bkg}).
\\ \indent Therefore a very low background is imperative to make a precise measurement. Scintillating cryogenic calorimeters provide a technique to reach very low background rates \cite{PirroScintBolo,DPoda2021,lumineu2017}. In particular, a scintillation light signal in coincidence with a heat signal in the calorimeter can be used to remove $\alpha$ particle backgrounds \cite{Cardani_LMO,lumineu2017,CUPID0_perf,LUCIFER_perf}.
\\ \indent In this letter we describe a measurement of the $2\nu\beta\beta$ decay rate and spectral shape of $^{100}$Mo using the CUPID-Mo experiment, a demonstrator for the next-generation $0\nu\beta\beta$ decay experiment CUPID \cite{CUPID_CDR}. A detailed description of the experiment can be found in \cite{CUPID_Mo_instrument}. It consisted of an array of 20 lithium molybdate (LMO) cryogenic calorimeters, enriched in $^{100}$Mo ($96.6\pm0.2$ \% isotope abundance) each of around 200 g mass. In addition, 20 germanium light detectors (LD), also operated as cryogenic calorimeters, were employed to readout the scintillation light signal used for particle identification to remove the $\alpha$ particle background. An individual module of CUPID-Mo consists of an LMO crystal attached to a copper holder and a Ge LD. 
Both the LMO and LD signals are readout using Neutron Transmutation Doped Germanium thermistors (NTD-Ge) \cite{Haller:1984}. These modules are then arranged into 5 towers of four LMOs each and was installed in the EDELWEISS cryostat \cite{EDELWEISS_perf} at the Laboratoire Soutterain de Modane (LSM), France. It collected a total exposure of $1.48 \ \mathrm{kg\times yr}$ of $^{100}$Mo between 2019 and 2020. The scintillation light signal allowed a complete rejection of $\alpha$ particles, while an excellent energy resolution of $7.7\pm 0.4$ keV FWHM was measured at 3034 keV \cite{CUPIDMo_0v}. This performance lead to a limit on $0\nu\beta\beta$ in $^{100}$Mo of $T^{0\nu}_{1/2}>1.8\times 10^{24}\, \mathrm{yr} \ \text{(90\% c.i.)}$ \cite{CUPIDMo_0v}.
\\ \indent
For this analysis we use the full data collected by CUPID-Mo. A detailed description of the data processing is given in \cite{CUPIDMo_0v} and was also used for \cite{ES,Bkg}. An optimal filter based analysis chain \cite{OF}, which maximizes the signal to noise ratio, is used to select physics events and estimate pulse amplitudes. Spurious events, such as pileup or spikes induced by electronics, are removed using a Principal Component Analysis based pulse shape cut \cite{PCA,CUPIDMo_0v}, normalized to ensure an energy independent efficiency. Due to the relatively short range of electrons in LMO, both $0\nu\beta\beta$ and $2\nu\beta\beta$ to ground states are likely to deposit energy in just a single LMO detector. However, background events induced by $\gamma$ quanta are more likely to deposit energy in multiple crystals. As such we define the {\it multiplicity} ($\mathcal{M}$) of an event as the number of LMO detectors with a pulse above the 40 keV energy threshold within a $\pm 10$ ms window.
 In addition, muon induced events are excluded using a dedicated muon veto system \cite{EDELWEISS:2013kzp}.
 We select $\beta$, $\gamma$ like events using the scintillation light signal as described in detail in \cite{CUPIDMo_0v}. We also remove events with a trigger in one LD with high $^{60}$Co contamination as described in \cite{Bkg,ES}. Multiplicity one $\gamma/\beta$ ($\mathcal{M}_{1,\gamma}/\beta$) events are used to extract the signal rates while $\mathcal{M}_2$ are used to constrain the $\gamma$ background. We also extract the spectra at high energy without any $\alpha$ rejection, this dataset ($\mathcal{M}_{1,\alpha}$) is used to constrain the radioactivity of the LMO crystals and other nearby components.
\\ \indent The energy resolution and bias in the energy scale are measured using $\gamma$ lines. The efficiency of all selection cuts has been estimated as $88.9\pm1.1$ \% for $\mathcal{M}_{1,\gamma/\beta}$ \cite{CUPIDMo_0v,Bkg}. No evidence of energy dependence was found, over the range of the fit, and our cuts are normalized to have an energy-independent efficiency.
\\ \indent
 To extract the rate of $2\nu\beta\beta$ decay we construct a model of the data described in detail in \cite{Bkg}. We simulate using GEANT4 \cite{Geant4} the $2\nu\beta\beta$ signal, using both the SSD and HSD models parameterised from \cite{kot_psf} and the contributions to the improved $2\nu\beta\beta$ model from Eq. \ref{imp} \cite{imp}. We also simulate radioactive contaminations in the various components of the experimental setup. These simulations are then convolved with a detector response model consisting of the energy resolution of the detectors, energy threshold, coincidences, and dead times of the detectors.
 \\ \indent We use a Bayesian analysis based on JAGS \cite{JAGS_1,JAGS_2} to fit our three experimental spectra ($\mathcal{M}_{1,\gamma/\beta},\mathcal{M}_{1,\alpha},\mathcal{M}_2)$ to a sum of MC simulations. The details of the choices of the model components are given in \cite{Bkg}.
 The fit to the $\mathcal{M}_{1,\gamma/\beta}$ spectrum uses a range of $100-4000$ keV. A variable binning  is used so that at minimum 15 keV bins are used in the continuum region. Each $\gamma$ or $\alpha$ peak is placed in one bin and then bins are combined so at-least 15 events are in each bin, to avoid the systematic effect of the peak line-shape.
 We show the $\mathcal{M}_{1,\gamma/\beta}$ fit in Fig. \ref{fig:fit}.
 We call this fit our {\it reference fit}: we see that this model is able to describe all the features of the experimental data, and that the data is dominated  by $2\nu\beta\beta$ decay events. While in \cite{Bkg} the SSD model of $2\nu\beta\beta$ was used by default, for this work we instead use the improved $2\nu\beta\beta$ model, and consider SSD as a cross check. By using the improved model, which allows the spectral shape to vary during the fit, we marginalize over the theoretical uncertainty in the spectral shape.
 \\ \indent 
 We study the consistency between our model and data using pseudo-experiments. We generate from the best fit model a set of 1000 pseudo-experiments, and for each we perform the background model fit and extract $-\log{(\mathcal{L})}$. The value obtained for the $\mathcal{M}_{1,\gamma/\beta}$ data is consistent with the expected distribution. In particular, we extract a p-value, or the probability of observing equal or larger fluctuations of $0.54$.
 \begin{table}[]
     \centering
          \caption{Systematic uncertainties in the determination of the $2\nu\beta\beta$ decay rate and $\xi_{3,1}$. All uncertainties are assigned either a Gaussian or asymmetric-Gaussian (for asymmetric uncertainties) posterior distribution with the exception of the $^{90}$Sr+$^{90}$Y where we assign a uniform distribution.}
     \begin{tabular}{c|c|c}
  Systematic test        & Uncert. $T_{1/2} [\%]$ & Uncert. $\xi_{3,1}$ [\%] \\ \hline \hline
   Source location      & $0.83$ & $0.9$\\
 $^{90}$Sr+$^{90}$Y & $+1.0$\footnote{Uniform distribution.} & $-4.9^{\text{a}}$ \\
  Minimal model & $0.24$ & $7.7$\\
  Binning & $0.37$ & $1.4$ \\
  Energy bias & $_{-0.16}^{+0.11}$ & $^{+3.5}_{-3.7}$\\
 Bremsstrahlung & $_{-0.22}^{+0.13}$ & $_{-6.8}^{+6.0}$\\
  MC statistics & $0.11$ & $1.4$\\
 Efficiency & $1.2$ & -  \\
 Isotopic abundance & $0.2$ & -
    
     \end{tabular}
     \label{tab:syst}
 \end{table}
 
\begin{figure}[h!]
    \centering
    \includegraphics[width=0.45\textwidth]{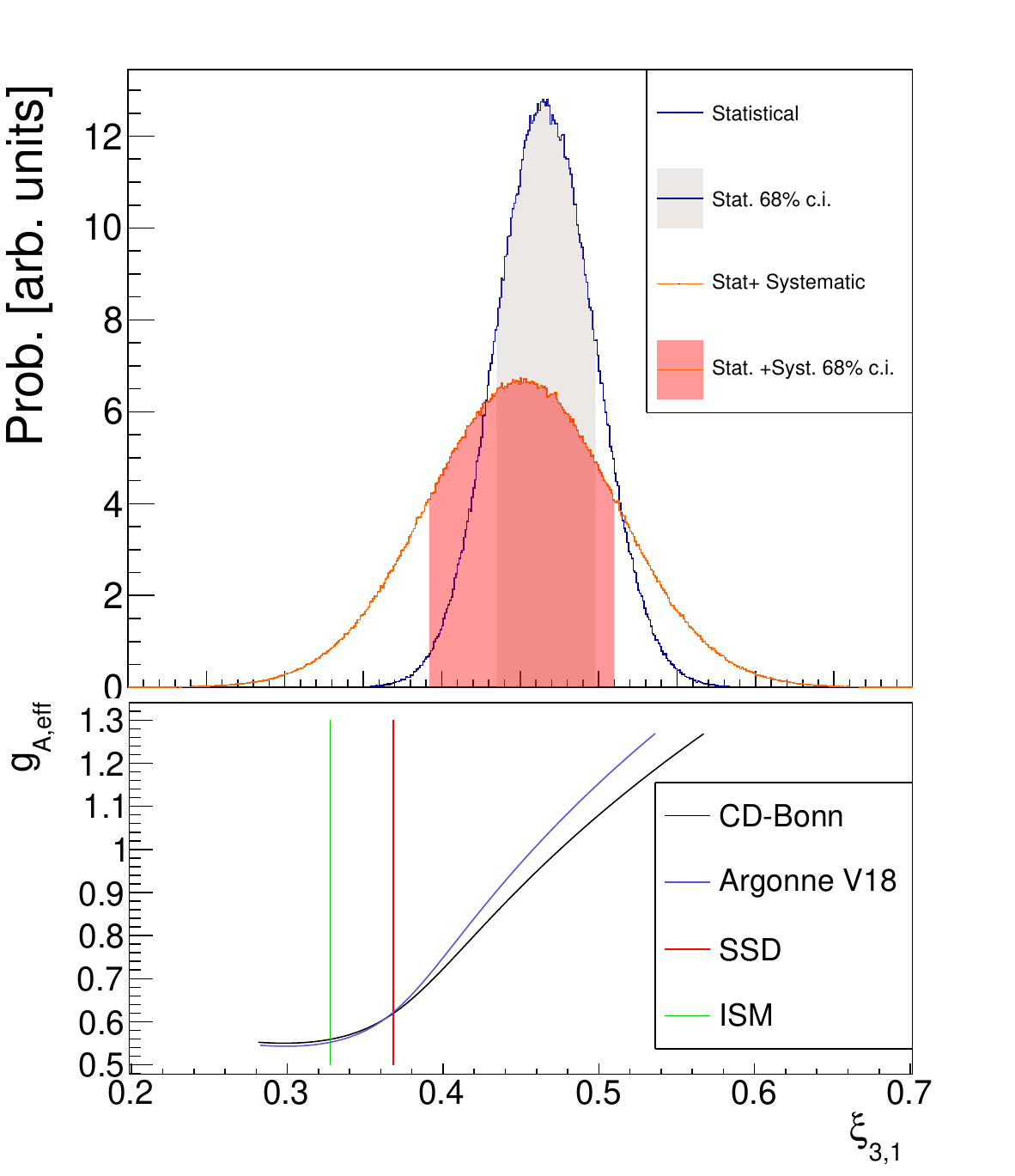}
    \caption{Posterior distribution of $\xi_{3,1}$ both with and without convolution with the systematic uncertainties (upper figure). In the lower panel we compare to the pn-QRPA, ISM and SSD theoretical values as a function of $g_{A,\text{eff}}$ for two potentials (CD-Bonn and Argonne V18 - see supplementary material for details). 
    }
    \label{xi_sum}
\end{figure}

From the background model fit we extract the $2\nu\beta\beta$ decay rate. 
We consider systematic uncertainties related to the number of reconstructed events and the efficiency and isotopic abundance conversion factors. 
We have performed a series of tests varying the assumptions of our background model to assess the dependence of $T_{1/2}$ on these choices. For each test a posterior distribution is assumed for the systematic uncertainty based on the change in the best fit value with respect to our reference fit. We then compute a convolution of these distributions and the posterior distribution from the fit to obtain the posterior distribution considering all systematic uncertainties. This can be considered a generalization of adding in quadrature to non-Gaussian uncertainties.
\\ \indent First we perform tests to check the dependence of our results on the $\gamma$ radioactivity source location.  We remove far sources of Th/U radioactivity leading to a slightly lower $2\nu\beta\beta$ rate ($-0.83\%$) and then  close (10 mK) sources of Th/U radioactivity leading to a higher rate ($+0.22\%$).
In principle, this uncertainty is already marginalized over in our analysis. However, our fit favours far sources of radioactivity, possibly due to some other effects such as pure $\beta$ decays, so we take a conservative approach considering an uncertainty of $\pm 0.83\%$ from the first test. This is assigned a Gaussian posterior to account for the possibility of even further sources than those included in our model.
\\ \indent In our background model we include a source of $^{90}$Sr+$^{90}$Y, consisting of two pure $\beta^-$ decays with Q-values $546$, $2276$ keV and around 60 hrs delay. In particular, the $^{90}$Y anti-correlates to the $2\nu\beta\beta$, in our model the activity is constrained as $179^{+36}_{-32} \ \mathrm{\mu Bq/kg}$. Again this uncertainty is in principle marginalized over already in our analysis. However, since the convergence of this parameter is driven by events at low energy we repeat the fit without the $^{90}$Sr+$^{90}$Y contribution. We obtain a half-life value $+1.0\%$ higher than the reference and we assign a uniform posterior between the reference and this fit.
\\ \indent We next repeat the fit removing any contributions where the smallest 68\% interval contains zero activity, which we call the {\it minimal model}. In our analysis, all the contributions are assigned non-negative uniform priors therefore a large number of parameters could bias the fit leading to a smaller $2\nu\beta\beta$ rate. We find a small shift of $+0.24\%$ in the $2\nu\beta\beta$ decay rate for this fit. We assign a Gaussian posterior distribution with $0.24$\% uncertainty for this systematic. We also check that our fit is not biased using our set of pseudo-experiments. The distribution of obtained $T_{1/2}$ is consistent with the fit to data.
\\ \indent Next we perform fits varying the energy scale by $\mp1$ keV resulting in a $2\nu\beta\beta$ decay rate shifted by $_{-0.16}^{+0.11}$\% which we assign an asymetric-Gaussian posterior.
\\ \indent Our reference fit uses a variable binning described in \cite{Bkg}. We repeat the fit using fixed binning of 1, 2, 10, 20 and 30 keV. The largest effect is for a binning of 2 keV where the rate is reduced by $-0.37\%$. We take a conservative approach considering a Gaussian posterior with $\pm 0.37\%$ standard deviation.
\\ \indent To assess the dependence on the accuracy of the MC simulations we generate simulations of $2\nu\beta\beta$ decay where we vary the Bremsstrahlung cross section by $\pm 10\%$. These lead to $_{-0.22}^{+0.13}$\% change in the $2\nu\beta\beta$ rate, which we assign an asymmetric-Gaussian posterior.
\\ \indent To account for the statistical uncertainty in the MC simulations we perform a fit adding nuisance parameters to the model as is done in \cite{CUPID_0_bkg}. This leads to a $-0.11\%$ smaller $2\nu\beta\beta$ rate, which we consider a systematic with a Gaussian posterior.
\\ \indent The final systematic uncertainties are on selection efficiency and $^{100}$Mo abundance which are $1.2\%$ and $0.2$\% respectively and are assigned Gaussian posteriors.
These systematic uncertainties are summarized in Tab. \ref{tab:syst}.
\\ \indent
Computing the convolution of all systematic uncertainties in Tab. \ref{tab:syst} and converting to the decay rate we compute the posterior distributions shown in Fig. \ref{post} for both the statistical only uncertainty and the combined uncertainty. From the central 68\% credible interval we extract a measurement of:
\begin{equation}
  T^{2\nu}_{1/2}= \meas \ \mathrm{yr}.
\end{equation}
With a relative uncertainty of $\pm 1.6$\% this is one the most precise determinations of a $2\nu\beta\beta$ decay half-life. The half-life is in agreement with our previous result obtained with a much smaller exposure \cite{lumineu_2vbb} and compatible with one obtained using the SSD $2\nu\beta\beta$ spectral shape model. 
\\ \indent
Next, we extract the values of the shape factors from the fit. We find that the parameters $\xi_{3,1}$ and $\xi_{5,1}$ are strongly anti-correlated ($\rho=-0.92$ - see more details in supplementary material).  $\xi_{5,1}/\xi_{3,1}$ is poorly constrained by our data with a best fit value of $1.6$ and a 90\% c.i. limit of $<40$. Within nuclear structure calculations the value of $\xi_{5,1}/\xi_{3,1}$ can be calculated reliably since $M_{GT-3}$ and $M_{GT-5}$ depend on contribution from low lying states. The value of $\xi_{5,1}/\xi_{3,1}$ within pn-QRPA is $0.364-0.368$ depending on $g_{A,\text{eff}}$ and the nuclear potential. Within the SSD hypothesis the value is $0.367$ \cite{imp} and within the ISM it is $0.349$ \cite{shell,shell2}. To reduce the degeneracy in our model we perform a fit with a Gaussian prior on $\xi_{5,1}/\xi_{3,1}$ with a mean of the SSD prediction and a conservative $5\%$ uncertainty.
\\ \indent
From this fit we extract the value of $\xi_{3,1}$ to compare to theoretical predictions. We consider the same systematic uncertainties as for the half-life (also shown in Tab. \ref{tab:syst}). The largest effects are found to be from the MC Bremstralhlung cross section and the choice of parameters of the model.
The posterior distribution of this observable both before and after convolution with the systematics is shown in Fig. \ref{xi_sum}. From this posterior distribution we extract a measurement:
\begin{equation}
   \xi_{3,1}=\ximeas.
\end{equation}
We compare our measurement of $\xi_{3,1}$ to pn-QRPA theoretical predictions in the lower panel of Fig. \ref{xi_sum}. Within pn-QRPA the $g_{pp}$ parameter (the strength of the particle-particle interaction) is tuned using our measurement of the half-life.
Since the calculated values of $\xi_{3,1}$ depend on $g_{A,\text{eff}}$, our measurement of $\xi_{3,1}$ provides complimentary information on $g_{A,\text{eff}}$. We find the experimental value is incompatible $(\sim 8\sigma$) with the prediction of the HSD hypothesis of $\xi_{3,1}=0$, somewhat incompatible with that from the ISM ($\sim 2.1\sigma$) but mostly compatible with that from the SSD hypothesis ($\sim 1.4\sigma$) and the pn-QRPA
predictions if the value of $g_{\text{A,eff}}$ is moderately quenched $(>0.8)$ or
unquenched. We encourage computation of $\xi_{3,1}$ and $\xi_{5,1}$
in additional theoretical frameworks such as the interacting Boson model \cite{Nomura:2022nwv}.
To extract a value for $g_{A,\text{eff}}$ within the pn-QRPA framework, we sample from the distribution of $\xi_{3,1}$  from our fit and for each sample we extract the corresponding $g_{A,\text{eff}}$ values. Assigning equal weights to the CD-Bonn and Argonne V-18 nuclear potientals we extract a posterior distribution on $g_{A,\text{eff}}$ and thus we extract a value:
\begin{equation}
    g_{A,\text{eff}}(\text{pn-QRPA})=1.0\pm 0.1 \ \text{(stat.)}\pm 0.2\text{(syst)}.
\end{equation}
As mentioned previously an analysis of $\xi_{3,1}$ and the half-life can be used to extract a measurement of $g_{A,\text{eff}}$ if $M_{GT-3}$ is known (see Eq. \ref{gA_form}).
Using the value of $M_{GT-3}$ from the ISM \cite{shell,shell2} and our fit we reconstruct:
\begin{equation}
    g_{A,\text{eff}}(\text{ISM})=\gAMeasSyst.
\end{equation}
The statistical uncertainty is obtained by sampling from the Markov chain, therefore combining the uncertainties on $T_{1/2}$ and $\xi_{3,1}$, the systematic uncertainty is obtained from the same tests as previously considered. This is the first measurement of $g_{A,\text{eff}}$ from a spectral shape study of $2\nu\beta\beta$ decay.
\\ \indent
In this letter we have reported a measurement of the $2\nu\beta\beta$ decay half-life of $^{100}$Mo. 
Utilizing excellent background rejection an almost background free spectrum is obtained which allows to obtain the most precise ever measurement of a $2\nu\beta\beta$ decay rate in this isotope.
Special attention was paid to the systematic uncertainties affecting the result, particularly to the source location, model choices and MC accuracy. 
\\ \indent In addition, we obtained a first of its kind measurement of a novel nuclear structure observable $\xi_{3,1}$ based on an improved description of the $2\nu\beta\beta$ decay process. The value of this observable is found to be incompatible with an HSD prediction, mildly incompatible with predictions from the ISM, but compatible with pn-QRPA predictions and a moderately quenched or unquenched value of $g_{A}^{\text{eff}}$. Finally, we report two novel measurements of $g_{A,\text{eff}}$, the first of their kind obtained from a spectral shape study of a $2\nu\beta\beta$ decay. 
\\ 
\indent This work has been performed in the framework of the CUPID-1 and   P2IO LabEx  programs, funded by the Agence Nationale de la Recherche (ANR, France). 
 F.A. Danevich, V.V. Kobychev, V.I. Tretyak and M.M. Zarytskyy were supported in part by the National Research Foundation of Ukraine.  A.S. Barabash, S.I. Konovalov, I.M. Makarov, V.N. Shlegel and V.I. Umatov were supported by the Russian Science Foundation. J. Kotila is supported by Academy of Finland.
Fedor ~\v{S}imkovic was partly supported by the Slovak Research and Development Agency.
Additionally the work is supported by the Istituto Nazionale di Fisica Nucleare (INFN), by the US Department of Energy (DOE) Office of Science.
This work makes use of the {\it Diana} data analysis software and the background model based on JAGS,  developed by the CUORICINO, CUORE, LUCIFER, and CUPID-0 Collaborations. Russian and Ukrainian scientists have given and give crucial contributions to CUPID-Mo. For this reason, the CUPID-Mo collaboration is particularly sensitive to the current situation in Ukraine. The position of the collaboration leadership on this matter, approved by majority, is expressed at \url{https://cupid-mo.mit.edu/collaboration#statement} . Majority of the work described here was completed before February 24, 2022.

\bibliography{bib.bib}
\clearpage
\appendix

\section{Supplementary material}
\begin{figure}
   \centering
  \includegraphics[width=0.45\textwidth]{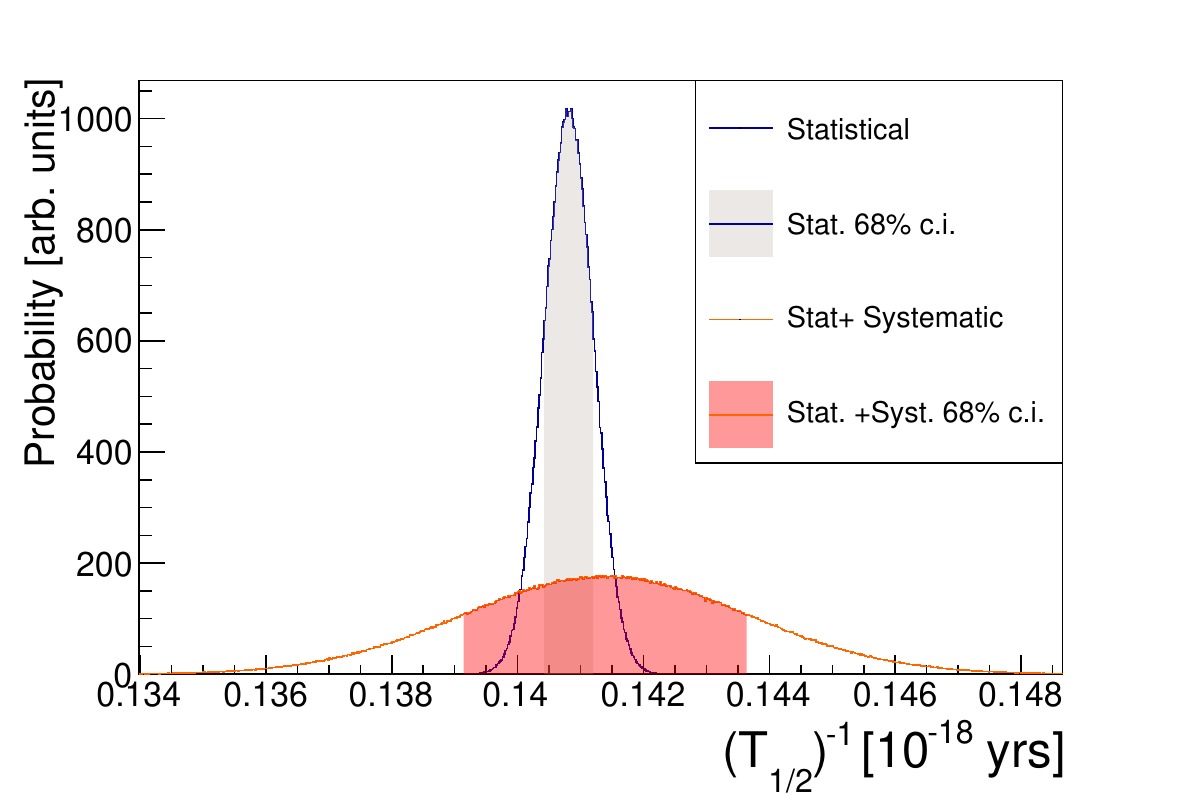}

 \caption{Posterior distribution on the inverse half-life $(T_{1/2})^{-1}$ for $2\nu\beta\beta$ decay of $^{100}$Mo both for the statistical only uncertainty and including all systematics.}
 \label{post}
\end{figure}
\begin{figure}
   \centering
   \includegraphics[width=0.45\textwidth]{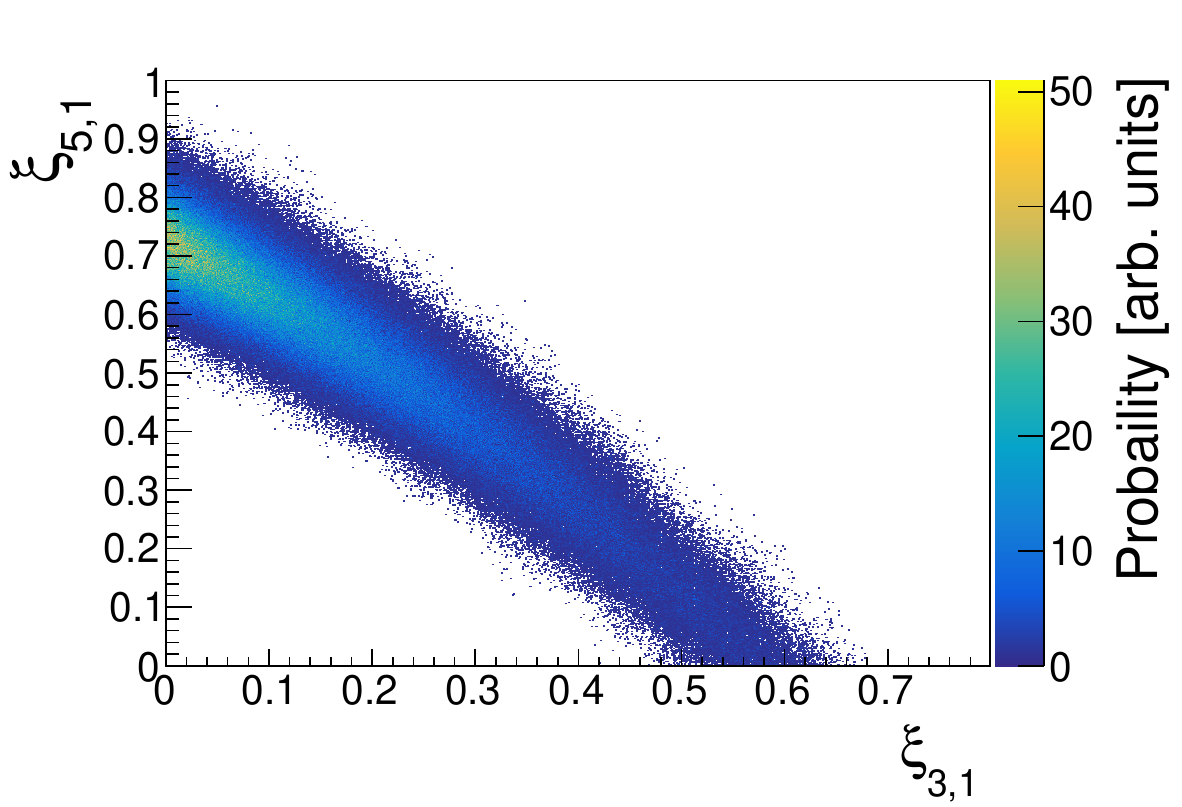}
 \caption{Two-dimensional posterior distribution of $\xi_{3,1}$ vs $\xi_{5,1}$. We observe a clear contribution from higher order terms and a strong but nonlinear correlation between $\xi_{3,1}$ and $\xi_{5,1}$.}
    \label{2D_xi}
\end{figure}
\subsection{QRPA theoretical predictions}
\label{theory}
Our pn-QRPA predictions were obtained considering the same large model space ($21$ subshells of N = $0–5$ oscillator shells) and mean fields as in \cite{Simkovic:2013qiy}. The pairing and residual interactions are derived from the same modern realistic nucleon-nucleon potentials, namely, from the charge-dependent Bonn potential (CD-Bonn) and the Argonne V18 potential. When solving the BCS pairing equations the strength of pairing interactions are slightly renormalized so that experimental pairing gaps are correctly reproduced.
The pn-QRPA equations contain three renormalization adjustable parameters: $g_{ph}$ for the particle-hole interaction, $g_{pp}^{T=1}$ and $g_{pp}^{T=0}$ for the isovector and isoscalar parts of the particle-particle interaction. While $g_{ph} = 1.0$ is typically used \cite{Simkovic:2013qiy}, $g_{pp}^{T=1}$ is fixed by the requirement that the $2\nu\beta\beta$ Fermi matrix element vanishes, as it should. It was found  $g^{T=1}_{pp} = 1.008$ $(0.9330)$ for Argonne (CD-Bonn) potential. $g^{T=0}_{pp}$ is adjusted so that the half-life of the $2\nu\beta\beta$ decay of $^{100}$Mo is correctly reproduced to the value previously measured for each considered value of $g_A^{\rm eff}$. Unlike earlier QRPA calculations of $2\nu\beta\beta$ NMEs the Bardeen-Cooper-Schrieffer (BCS) overlap of the initial and final vacua is taken into account leading to a reduction by factor $0.87$ \cite{Simkovic:2003rk}. $g_A^{\rm eff}=$ 0.8, 1.00 and 1.27 was reproduced with $g^{T=0}_{pp} =$ 0.817 (0.746), 0.837 (0.767) and 0.846 (0.776) for Argonne (CD-Bonn) potential, respectively.

\subsection{Posterior distributions}
\label{correlation}
\begin{figure}
   \centering
 \includegraphics[width=0.49\textwidth]{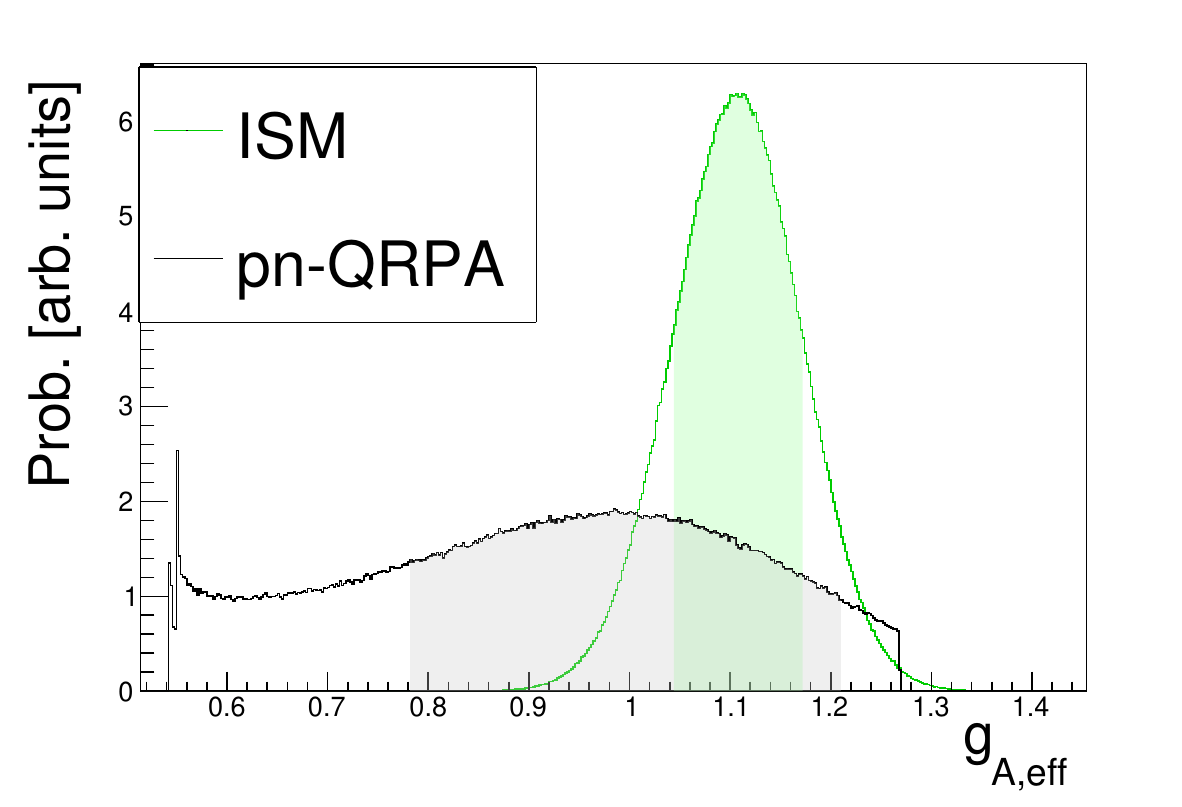}
 \caption{Posterior distribution on $g_{A,\text{eff}}$ base on the analysis within the pn-QRPA and the ISM frameworks. For the pn-QRPA analysis, the posterior features a sharp peak around $g_{A,\text{eff}}\sim 0.55$ due to the non linear relationship between $g_{A,\text{eff}}$ and $\xi_{3,1}$. 
   }
   \label{gA_post}
\end{figure}

We provide for reference the posterior distribution on the inverse half-life of $2\nu\beta\beta$ (in Fig. \ref{post}), a two dimensional posterior distributions showing the correlation between the various observables of the spectral shape $\xi_{3,1}$ against $\xi_{5,1}$ (in Fig. \ref{2D_xi}) and posterior distributions on $g_{A,\text{eff}}$ for the analysis within the ISM and pn-QRPA (Fig. \ref{gA_post}).

\end{document}